\documentclass[aps,prc,twocolumn]{revtex4-1}
\usepackage{graphicx}
\usepackage{amsmath,amssymb}

\begin{document}
\title{Pressure Anisotropy in Heavy Ion Collisions from Color Glass Condensate}
\date{\today}
\author{Ming Li}
\author{Joseph I. Kapusta}
\affiliation{School of Physics and Astronomy, University of Minnesota,
Minneapolis MN 55455, USA}
\begin{abstract}
We generalize calculations of the energy-momentum tensor for classical gluon fields in the McLerran-Venugopalan model using the small-$\tau$ power series expansion method. Results to all orders for the energy density and pressures are given in the leading $Q^2$ approximation and with the inclusion of running coupling effects.  The energy density and transverse pressure decrease monotonically with time while the longitudinal pressure starts from a negative value and increases towards zero.
\end{abstract}

\maketitle

\section{Introduction}

It is well known that the quark-gluon plasma (QGP) produced in heavy ion collisions can be successfully described by hydrodynamics; successful hydrodynamics requires rapid thermalization with the thermalization time $\tau_0$ less than 1 fm/c \cite{Heinz2003}. Studying the dynamical processes before the formation of the QGP can provide initial conditions for the hydrodynamic simulations.  During the hydrodynamic evolution, the system is assumed to be close to local thermal equilibrium so that the use of  thermodynamic quantities and relations are justified. On the other hand, ideal hydrodynamics implies that the system is isotropic in the local rest frame while viscous hydrodynamics could accomodate a small amount of anisotropy. Considering the initial conditions, a natural question to ask is whether the system becomes isotropic at or before the thermalization time $\tau_0$. The question is significant in the sense that isotropization is closely related to thermalization. Study of the isotropization problem can be beneficial to the understanding of the early thermalization puzzle. Furthermore, unraveling the isotropization process can determine whether we need isotropic or anisotropic initial conditions for the subsequent hydrodynamic evolutions. The problem has been explored by several authors. Chesler and Yaffe \cite{Chesler:1,Chesler:2} and Heller, Janik and Witaszczyk \cite{Heller:2011ju} approached the problem by studying the strongly coupled plasma in the $\mathcal{N}=4$ supersymmetric gauge theory under the AdS/CFT correspondence.  They found sizable anisotropy in the longitudinal and transverse pressures $P_L$ and $P_T$ when viscous hydrodynamics is presumed to become applicable.  Epelbaum and Gelis \cite{Gelis:2013rba} numerically solved the classical SU(2) Yang-Mills equations on a lattice in the Color Glass Condensate (CGC) framework.  They found that $P_T/\varepsilon \approx 0.5$ and $P_L/\varepsilon \approx 0$, where $\varepsilon$ is the energy density, when $Q_s \tau > 1$.  When initial quantum fluctuations were included, and using $g = 0.5$, they found a pressure anisoptropy $P_L/P_T\approx 0.6$.  Finally, Strickland \cite{Strickland:2014pga} argued for the pressure anisotropy in the QGP from the viscous hydrodynamics itself and proposed an anisotropic hydrodynamics. 

In this paper, we will address the pressure isotropization problem in the framework of the CGC \cite{McLerran:1993ni,McLerran:1993ka}. Instead of numerically solving the classical Yang-Mills equations, we use the small-$\tau$ power series expansion method \cite{Fries:2006pv} to analytically solve the  equations. To be specific, we generalize the calculations of the energy-momentum tensor of the classcial gluon fields in \cite{Chen:2015wia} to all orders in $\tau$. We also include the running coupling constant effects.  

The paper is organized as follows. In section \ref{cgc_smallt}, we briefly review the CGC framework and the application to heavy ion collisions. Then we discuss the small-$\tau$ power series expansion method and set up the calculations for the energy-momentum tensor. In section \ref{em_tensor}, we present the all-order calculations with the leading $Q^2$ approximation and explain how we include running coupling constant effects. In section \ref{numerical}, numerical calculations are given with discussions of the physics implied.  Section \ref{conclusion} summarizes the results.  Technical details of the calculational steps are given in the Appendix. 

\section{Color Glass Condensate Framework}\label{cgc_smallt}

\subsection{The McLerran-Venugopalan (MV) Model}

In the CGC framework \cite{Iancu:2002xk,Iancu:2003xm}, the high energy limit of the hadron wavefunctions are approximated by the small-x partons while the large-x partons serve as the color sources radiating small-x partons. The small-x partons are saturated with typical transverse momentum $Q_s$, which depends on the collison energy (thus x) and the size of the colliding nuclei. Due to the overpopulated nature of the small-x partons in each momentum mode, a classical description using the classical gluon fields $A^{\mu}(x)$ is sufficient to describe the small-x partons. On the other hand, the color sources are traveling at the speed of light and ignore the back-reaction of the classical field on the sources. Also, the time hierarchy of the bremsstrahlung ladder implies that the color sources are static and randomly distributed.  Mathematically, it is equivalent to the classical Yang-Mills equation
\begin{equation}\label{ym_eq}
[D_{\mu},F^{\mu\nu}] = j^{v} \,,
\end{equation}
with $j^{v} = \delta^{\nu +} \rho(x^{-}, \vec{x}_{\perp}) $. Here we used the light-cone coordinates $x^+ = (t+z)/\sqrt{2}, \,x^- =(t-z)/\sqrt{2}$. The color charge density $\rho(x^-,\vec{x}_{\perp})$ is a random (field) variable. It obeys the probability distribution $W_{\Lambda}[\rho]$ depending on the scale $\Lambda$ separating small-x and large-x. The probability distribution $W_{\Lambda}[\rho]$ follows the renormalization group equation, the JIMWLK equation \cite{JalilianMarian:1996xn,JalilianMarian:1997gr, JalilianMarian:1997dw,Iancu:2000hn,Ferreiro:2001qy} which is in general difficult to solve.  In the classical McLerran-Venugopalan model \cite{McLerran:1993ni,McLerran:1993ka}, the color charge sources are assumed to be independent and uncorrelated and they satisfy Gaussian distributions. The same information is encapsulated in the two-point correlation function \cite{Chen:2015wia},
\begin{equation}\label{color_charge_correlation}
\begin{split}
&\langle \rho_a(x^-,\vec{x}_{\perp})\rho_b(y^-,\vec{y}_{\perp})\rangle \\
=& \frac{g^2}{d_A}\delta_{ab}\lambda(x^-,\vec{x}_{\perp})\delta(x^--y^-)
\delta^{(2)}(\vec{x}_{\perp}-\vec{y}_{\perp}) \, . \\
\end{split}
\end{equation}
Here $d_A = N^2_c-1$ is the dimension of the adjoint representation of the $\rm{SU}(\rm{N}_c)$ group and $a,b$ are color indexes. The function $\lambda(x^-,\vec{x}_{\perp})$ appears in the integral $\int dx^- \lambda(x^-,\vec{x}_{\perp}) = \mu(\vec{x}_{\perp})$ where $\mu(\vec{x}_{\perp})$ has the meaning of color charge squared per unit area. Notice that we adopt a different normalization in Eq. \eqref{color_charge_correlation} compared with those given in \cite{Iancu:2002xk,Iancu:2003xm}. 

In applications to heavy ion collisions \cite{Kovner:1995ts,Kovner:1995ja}, the color current $j^{\mu}$ has two parts coming from the two colliding nuclei: $j^{\mu} = j^{\mu}_1 + j^{\mu}_2=\delta^{\mu +}\rho_1(x^-,\vec{x}_{\perp}) + \delta^{\mu -}\rho_2(x^+,\vec{x}_{\perp})$. One nucleus travels along the forward light cone $x^+$ while the other one travels along the backward light cone $x^-$.  We are interested in the dynamics after the collision, that is, in the space-time region of $x^+>0,\, x^->0$. In this region, the classical Yang-Mills equations \eqref{ym_eq} are sourceless, $j^{\nu} =0$. If we further choose the Fock-Schwinger gauge $\tau A^{\tau} =x^+A^-+x^-A^+=0$ and assume boost-invariant solutions $A^{\eta}(\tau,\vec{x}_{\perp})$ and $A^i_{\perp}(\tau,\vec{x}_{\perp})$, Eq.\eqref{ym_eq} can be explicitly written as the equations of motion
\begin{equation}\label{eoms}
\begin{split}
&\frac{1}{\tau}\frac{\partial}{\partial \tau}\frac{1}{\tau}\frac{\partial}{\partial \tau} \tau^2 A^{\eta} - [D^i,[D^i,A^{\eta}]] =0 \, , \\
&\frac{1}{\tau}\frac{\partial}{\partial \tau} \tau \frac{\partial}{\partial \tau} A_{\perp}^i -ig\tau^2 [A^{\eta},[D^i,A^{\eta}]] - [D^j,F^{ji}] =0 \, ,\\
\end{split}
\end{equation}
and the constraint equation
\begin{equation}\label{constraint}
ig\tau [A^{\eta},\frac{\partial}{\partial \tau} A^{\eta}] - \frac{1}{\tau} [D^i,\frac{\partial}{\partial \tau} A^i_{\perp}] =0 \,.
\end{equation}
In the above expresions, we changed the coordinate system from the light-cone coordinates $(x^+,x^-,\vec{x}_{\perp})$ to the Milne coordinates $(\tau,\eta,\vec{x}_{\perp})$. They are related by $\tau =\sqrt{2x^+x^-}$ and $\eta=\frac{1}{2}\ln (x^+/x^-)$. In the rest of the paper, we will use the Milne coordinates exclusively.  The dynamics before the collision, that involve the space-time regions $x^+>0,\,x^-<0$ and $ x^+<0,\,x^->0$, provide the necessary initial conditions \cite{Kovner:1995ja, Gyulassy:1997vt} for the equations of motion \eqref{eoms}
\begin{equation}\label{ics}
\begin{split}
&A^i_{\perp}(\tau=0,\vec{x}_{\perp}) = A_1^i(\vec{x}_{\perp}) + A^i_2(\vec{x}_{\perp}) \, , \\
&A^{\eta}(\tau=0,\vec{x}_{\perp}) = -\frac{ig}{2}[A_1^i(\vec{x}_{\perp},A_2^i(\vec{x}_{\perp}] \, , \\
&\frac{\partial}{\partial\tau} A^i_{\perp}(\tau=0,\vec{x}_{\perp}) =0 ,\quad \frac{\partial}{\partial\tau} A^{\eta} (\tau=0,\vec{x}_{\perp}) =0 \,.
\end{split}
\end{equation}
Here $A^i_1(\vec{x}_{\perp})$ and $A^i_2(\vec{x}_{\perp})$ are the pure gauge fields produced by the single-nucleus color sources $\rho_1$ and $\rho_2$, respectively.  

In summary, study of the post impact dynamics of heavy ion collisions in the MV model is equivalent to solving the initial value problem \eqref{eoms}-\eqref{ics}. Event averaged properties are obtained afterwards by taking the statistical average through the two-point correlation function \eqref{color_charge_correlation}. 

\subsection{Small-$\tau$ Expansion}

The initial value problem \eqref{eoms}-\eqref{ics} has been studied analytically in \cite{Kovner:1995ts,Kovner:1995ja} where the color sources were assumed weak so that perturbative expansion in terms of the color sources $\rho$ is justified. Another analytical approach proposed in \cite{Fries:2006pv,Chen:2015wia} solves the classical Yang-Mills equation by the power series expansion in proper time $\tau$. This approach is valid as long as we focus on early time (small $\tau$) within the convergence radius set by $1/Q_s$ and  there is no initial singularity at $\tau=0$ .  There are no generic analytic solutions although numerical solutions have been investigated by several groups \cite{Krasnitz:1999wc,Krasnitz:2001qu,Lappi:2003bi,Lappi:2006hq}. We will follow the small-$\tau$ power series expansion method. The solutions are represented by
\begin{equation}
\begin{split}
&A^\eta(\tau,\vec{x}_{\perp}) =\sum_{n=0}^{\infty}\tau^n A^\eta_{(n)}(\vec{x}_{\perp}) \, , \\
&A^i_{\perp}(\tau,\vec{x}_{\perp}) = \sum_{n=0}^{\infty}\tau^n A_{\perp (n)}^i(\vec{x}_{\perp}) \, .\\
\end{split}
\end{equation}
Substituting into Eqs. \eqref{eoms}, one can check that coefficients with odd n vanish and that we obtain the following recursion relations $(n\geq1)$
\begin{equation}\label{recursive_A}
\begin{split}
&A^\eta_{(2n)} = \frac{1}{2n(2n+2)} \sum_{k+l+m=2n-2}\left[D^i_{(k)},\left[D^i_{(l)},A^\eta_{(m)}\right]\right] \, , \\
&A^i_{(2n)} = \frac{1}{(2n)^2}\Big(\sum_{k+l=2n-2}\left[D^j_{(k)},F^{ji}_{(l)}\right] \\
&\qquad\qquad\qquad + ig\sum_{k+l+m=2n-4} \left[A^\eta_{(k)},\left[D^i_{(l)},A^\eta_{(m)}\right]\right]\Big) \, . \\
\end{split}
\end{equation}
With the zeroth order coefficients given in \eqref{ics}, we can solve $A^\eta_{(2n)}$ and $A^i_{\perp(2n)}$ order by order. Furthermore, the field strength tensor $F^{\mu\nu}$ can be obtained as a power series expansion of $\tau$ by
\begin{equation}
\begin{split}
&F^{\tau\eta} =\partial^{\tau}A^{\eta} \, , \qquad F^{\tau i} =\partial^{\tau}A^i,\qquad F^{\eta i} = -[D^i,A^\eta] \, , \\
&F^{ij}=\partial^i A^j-\partial^jA^i-ig[A^i,A^j] \, .
\end{split}
\end{equation}
Denote $\tilde{F}^{\tau\eta} \equiv \tau F^{\tau\eta}$ and $\tilde{F}^{\eta i} \equiv \tau F^{\eta i}$ to avoid the coordinate singularities. The recursion relations for the field strength tensor components  are
\begin{align}\label{recursive_F}
&\tilde{F}^{\tau\eta}_{(2n)} = (2n+2) A^{\eta}_{(2n)} \nonumber \, , \\
&F^{\tau i}_{(2n-1)}=(2n) A^{i}_{(2n)} \, , \nonumber\\
&\tilde{F}^{\eta i}_{(2n-1)} = - \left[D^i_{(0)},A^{\eta}_{(2n-2)}\right] + ig \sum_{\substack{k+l=2n-2\\ k\neq 0}} \left[A^{i}_{(k)},A^{\eta}_{(l)}\right] \, , \nonumber \\
&F^{ij}_{(2n-2)} = \left[D^{i}_{(0)},A^j_{(2n-2)}\right] - \left[D^{j} _{(0)} ,A^i_{(2n-2)}\right]  \nonumber\\
& \quad\qquad\qquad-ig\sum_{\substack{k+l=2n-2\\k\neq 0,\,l\neq 0}}\left[A^i_{(k)},A^j_{(l)}\right] \, . 
\end{align}
At zeroth order, that is when $\tau=0$, the only nonvanishing components of the field strength tensor are the longitudinal chromo-electric field $E_0$ and longitudinal chromo-magnetic field $B_0$, specifically
\begin{equation}
\begin{split}
&E_0 = -\tilde{F}^{\tau\eta}_{(0)} = ig\delta^{ij}[A_1^i,A_2^j] \, , \\
&B_0 = -F^{12}=ig\epsilon^{ij}[A^i_1,A^j_2] \, . \\
\end{split}
\end{equation}
Here $\delta^{ij}$ and $\epsilon^{ij}$ are the two-dimensional Kronecker delta function and Levi-Civita symbol, respectively. 

\subsection{Energy-Momentum Tensor of The Glasma}
The classical field solutions obtained in the above subsection describe the overoccupied off-shell soft gluon system. The field is stong in the sense that $A\sim 1/g$. This state of matter created right after a heavy-ion collision is dubbed Glasma \cite{Lappi:2006fp}. The question of how gluons are liberated as on-shell particles from the Glasma, as well as the production of quarks and their thermalization, remains an open question. Here, we will content ourselves with one of the properties of the Glasma state---the event averaged energy-momentum tensor $T^{\mu\nu}$---and explore its spacetime evolution.

The energy-momentum tensor is
 \begin{equation}
 T^{\mu\nu} = F^{\mu\lambda}F_{\,\, \lambda}^{\nu} + \frac{1}{4}g^{\mu\nu}F^{\kappa\lambda}F_{\kappa\lambda} \, ,
 \end{equation} 
with a trace over color indices taken implicitly as $AB=2\,\mbox{Tr}(AB)$. Assuming boost-invariance, the energy-momentum tensor can be parameterized as
\begin{widetext}
\begin{equation}\label{emtensor}
T^{\mu\nu}=
\begin{pmatrix}
\mathcal{A}+\mathcal{B}\cosh{2\eta}+\mathcal{C}\sinh{2\eta} & \mathcal{E}_1\sinh{\eta}+\mathcal{F}_1\cosh{\eta} & \mathcal{E}_2\sinh{\eta}+\mathcal{F}_2\cosh{\eta} & \mathcal{B}\sinh{2\eta}+\mathcal{C}\cosh{2\eta} \\
\mathcal{E}_1\sinh{\eta}+\mathcal{F}_1\cosh{\eta} & \mathcal{A}+\mathcal{D} & \mathcal{G} & \mathcal{F}_1\sinh{\eta} +\mathcal{E}_1\cosh{\eta} \\
\mathcal{E}_2\sinh{\eta}+\mathcal{F}_2\cosh{\eta} & \mathcal{G} & \mathcal{A}-\mathcal{D} & \mathcal{F}_2\sinh{\eta} +\mathcal{E}_2\cosh{\eta} \\
\mathcal{B}\sinh{2\eta}+\mathcal{C}\cosh{2\eta} & \mathcal{F}_1\sinh{\eta} +\mathcal{E}_1\cosh{\eta} & \mathcal{F}_2\sinh{\eta} +\mathcal{E}_2\cosh{\eta} & -\mathcal{A}+\mathcal{B}\cosh{2\eta}+\mathcal{C}\sinh{2\eta}\\
\end{pmatrix}
\end{equation}
\end{widetext}
where $\mathcal{A}, \mathcal{B}, \mathcal{C}, \mathcal{D}$, $\vec{\mathcal{E}} =(\mathcal{E}_1,\mathcal{E}_2)$ , $\vec{\mathcal{F}}=(\mathcal{F}_1,\mathcal{F}_2)$ , $\mathcal{G}$ are functions of proper time $\tau$ and transverse spatial coordinate $\vec{x}_{\perp}$. Explicit expressions are
\begin{equation}
\begin{split}
&\mathcal{A} = \frac{1}{2}\left(\tilde{F}^{\tau\eta}\tilde{F}^{\tau\eta} +\frac{1}{2} F^{ij}F^{ij}\right) \, ,\\
& \mathcal{B} = \frac{1}{2}(F^{i\tau}F^{i\tau}+\tilde{F}^{i\eta}\tilde{F}^{i\eta}) \, , \quad \mathcal{C} = F^{i\tau}\tilde{F}^{i\eta} \, , \\
& \mathcal{D} = -\frac{1}{2} [F^{x\tau}F^{x\tau}-F^{y\tau}F^{y\tau}-(\tilde{F}^{x\eta}\tilde{F}^{x\eta}-\tilde{F}^{y\eta}\tilde{F}^{y\eta})] \, ,\\
&\mathcal{E}^i = F^{i\tau}\tilde{F}^{\tau\eta} -F^{ij}\tilde{F}^{j\eta},\qquad \mathcal{F}^i =\tilde{F}^{i\eta}\tilde{F}^{\tau\eta} - F^{ij}F^{j\tau} \, , \\
&\mathcal{G} = -F^{x\tau}F^{y\tau} + \tilde{F}^{x\eta}\tilde{F}^{y\eta} \, .
\end{split}
\end{equation}
In the general situation where the color charge fluctuation $\mu(\vec{x}_{\perp}) = \int dx^- \lambda(x^-,\vec{x}_{\perp})$ depends on the transverse coordinates $\vec{x}_{\perp}$, the energy-momentum tensor was solved up to fourth order in $\tau$ and first order in the gradients of $\mu(\vec{x}_{\perp})$ \cite{Chen:2015wia}. At zeroth order, the initial energy-momentum tensor is diagonal $T^{\mu\nu}_{(0)} = \mbox{diag}\{\varepsilon_0,\varepsilon_0,\varepsilon_0,-\varepsilon_0\}$ with
\begin{equation}
\label{initial_energy}
\varepsilon_0(\vec{x}_{\perp}) = 2\pi\alpha_s^3\frac{C_A}{d_A}\mu_1(\vec{x}_{\perp})\mu_2(\vec{x}_{\perp})\ln\left(\frac{Q^2_1}{m^2_1}\right)\ln\left(\frac{Q^2_2}{m^2_2}\right) \, .
\end{equation}
Here $C_A=N_c$ is the Casmir operator of $\rm{SU}(\rm{N}_c)$ in the adjoint representation and $d_A = N_c^2-1$. The $Q_i$ and $m_i$ $(i=1,2)$ are the UV scale and IR scale of soft gluon modes for the two nuclei. For the nonequilibrium state of the gluon fields, we will denote the longitudinal pressure as $P_L = T^{33}$ and the transverse pressure as $P_T=(T^{11}+T^{22})/2$. With these definitions, one can see the highly anisotropic nature of the initial pressures $P_T =\varepsilon_0$ and $P_L=-\varepsilon_0$ and their relatively large value compared to a thermalized system, like $P=\varepsilon/3$ for a relativistic gas. 

\section{Calculation of the Energy-Momentum Tensor to all Orders}
 \label{em_tensor}

 \subsection{General Expressions}

In this section we focus on a simplified problem where the transverse color charge fluctuations are homogeneous and isotropic $ \mu(\vec{x}_{\perp})=\mu$ and the two colliding nuclei are the same $\mu_1=\mu_2$. The physical quantities to be considered are energy density and pressure as functions of proper time. The early time behavior of the energy density is crucial for understanding the inital conditions for hydrodynamics. The evolution of transverse pressure and longitudinal pressure can demonstrate the possible isotropization processes.
 
Under the condition of transverse homogeneity and isotropy, the energy-momentum tensor \eqref{emtensor} reduces to
\begin{equation}\label{emtensor2}
T^{\mu\nu}=
\begin{pmatrix}
\mathcal{A}+\mathcal{B}\cosh{2\eta} & 0 & 0 & \mathcal{B}\sinh{2\eta} \\
0 & \mathcal{A} & 0 & 0 \\
0 & 0 & \mathcal{A} & 0 \\
\mathcal{B}\sinh{2\eta} & 0 & 0 & -\mathcal{A}+\mathcal{B}\cosh{2\eta}\\
\end{pmatrix} \, .
\end{equation}
The energy and momentum of the glasma itself are conserved as we assume the two receding nuclei still propagate at the speed of light and ignore the back-reaction of the classical field on the nuclei. Implementing $\partial_{\mu}T^{\mu\nu} =0$, we obtain the relation between $\mathcal{A}$ and $\mathcal{B}$
\begin{equation}
\frac{\partial}{\partial \tau}(\tau^2\mathcal{B}) +\tau^2\frac{\partial \mathcal{A}}{\partial \tau} =0 \, .
\end{equation}
Plugging into the power series expansion of $\mathcal{A}(\tau)$ and $\mathcal{B}(\tau)$
\begin{equation}\label{aandb}
\mathcal{A}(\tau) = \sum_{n=0}^{\infty}\tau^{2n} \mathcal{A}_{(2n)},\quad \mathcal{B}(\tau) = \sum_{n=0}^{\infty}\tau^{2n} \mathcal{B}_{(2n)} \, ,
\end{equation}
we get the order-by-order relations 
\begin{equation}\label{aandb2}
\mathcal{B}_{(2n)} = -\frac{2n}{2n+2} \mathcal{A}_{(2n)} \, .
\end{equation}
Therefore, all we need to calculate are the coefficients $\mathcal{A}_{(2n)}$.

The general expression for $\mathcal{A}_{(2n)}$ is
\begin{equation}\label{a2n_cal}
\begin{split}
\mathcal{A}_{(2n)} & \sim [D^{i_1},[D^{i_2},[D^{i_3},\ldots,[D^{i_{n}},[A^p_1,A^q_2]]\ldots,]]]\\
&\times [D^{j_1},[D^{j_2},[D^{j_3},\ldots,[D^{j_{n}},[A_1^m,A^n_2]\ldots,]]] \, , \\
\end{split}
\end{equation}
with event averaging using Eq. \eqref{color_charge_correlation} in mind. There is a complicated overall prefactor that contracts with the spatial indexes $i_1,i_2,\ldots i_{n}; j_1,j_2,\ldots,j_{n};$ $ m,n,p,q;$ so that the final expression for $\mathcal{A}_{(2n)}$ is index free. It affects the coefficient of each term but will not influence the general structure of each term. In the expression for $\mathcal{A}_{(2n)}$, there are $2n$ covariant derivatives $D^i$ which can be written in different ways: $D^i \equiv\partial^i - igA^i_1-igA^i_2 = D^i_1-igA^i_2 = D^i_2-igA^i_1$. We organize all terms contained in $\mathcal{A}_{(2n)}$ according to the number of covariant derivatives involved. Henceforth we employ the following correlation functions (we use $A_1^i$ as an example, similar results hold for $A_2^i$). 
\begin{equation}\label{appoximation1}
\begin{split}
&\langle D^{i_1}_1D^{i_2}_1\ldots D^{i_k}_1A_1^pA_1^m\rangle \sim \frac{g^2}{d_A}\left(\frac{\mu}{4\pi}\right) Q^k \, , \\
&\langle A_1^pA_1^m\rangle \sim \frac{g^2}{d_A}\left(\frac{\mu}{4\pi}\right) \ln\left(\frac{Q^2}{m^2}\right) \, , \\
&\langle A_1^{i_1}A_1^{i_2}\ldots A_1^{i_k} \rangle \sim\sum_{\mbox{all combinations}} \langle A_1^{i_{p1}}A_1^{i_{p2}}\rangle \langle A_1^{i_{p3}}A_1^{i_{p4}}\rangle \ldots
\end{split}
\end{equation}
(Formulas with the coefficients included are given in Appendix \ref{appendixA}).  Here $Q$ is a UV cut-off in transverse momentum space as can be seen from the explicit calculation of the correlation functions in Eq. \eqref{uv_calculate}. The $m$ an IR cut-off regulating the low energy behavior. We assume $Q^2\gg m^2$ so that we only need to keep $(Q^2)^k$ terms and disregard $(m^2)^k\ln (Q^2/m^2)$ and $(Q^2)^l(m^2)^{k-l}$ terms in the calculation of the correlation functions. We further used the fact that only two-point correlations of color sources are nonvanishing \cite{Kovchegov:1996ty,Fukushima:2007dy}

After calculating the first few orders, the general expression for $\mathcal{A}_{(2n)}$ can be parameterized as
\begin{equation}\label{a2n}
\begin{split}
&\mathcal{A}_{(2n)} = g^6\frac{C_A}{d_A}\left(\frac{\mu}{4\pi}\right)^2\ln\left(\frac{Q^2}{m^2}\right) \\
&\times \sum_{k=1}^{n}  \mathsf{F}_{2n}(k) \left(Q^2\right)^k \left[g^4\frac{C_A}{d_A}\frac{\mu}{4\pi} \ln\left(\frac{Q^2}{m^2}\right)\right]^{n-k}\\
&+g^6\frac{C_A}{d_A}\left(\frac{\mu}{4\pi}\right)^2\sum_{k=2}^n \mathsf{G}_{2n}(k) \left(Q^2\right)^k \left[g^4\frac{C_A}{d_A}\frac{\mu}{4\pi}\ln\left(\frac{Q^2}{m^2}\right)\right]^{n-k}\\
& +g^6\frac{C_A}{d_A}\left(\frac{\mu}{4\pi}\right)^2\ln^2\left(\frac{Q^2}{m^2}\right)  \left[g^4\frac{C_A}{d_A}\frac{\mu}{4\pi}\ln\left(\frac{Q^2}{m^2}\right) \right]^n \mathsf{H}_{2n} \, .
\end{split}
\end{equation}
At each step, the number of covariant derivatives are reduced by two, and we have one additional factor $g^2 A_1A_1$. That is why $Q^2$ is replaced by 
$g^4 C_A \mu/4\pi d_A\ln (Q^2/m^2)$ when its power index is descending.  All the numerical coefficients $\mathsf{F}_{2n}(k),\mathsf{G}_{2n}(k),\mathsf{H}_{2n}$ have to be determined by detailed calculations which, in general, are hard to achieve.

\subsection{The Leading $Q^2$ Approximation and Running Coupling Effects}

Within the CGC framework, the saturation scale $Q_s$ is related to the color charge fluctuation measure $\mu$ self-consistently by \cite{Iancu:2002xk}
\begin{equation}\label{mu_Qs}
Q_s^2 = g^4(Q_s^2)\frac{C_A}{d_A} \frac{\mu}{16\pi}\ln\left(\frac{Q_s^2}{m^2}\right) \, .
\end{equation}
Here we explicitly write out the energy scale dependence of the strong coupling constant $g(Q_s^2)$. The lowest-order perturbative calculation of the running coupling constant is
\begin{equation}\label{running_coupling}
\alpha_s(M^2) = \frac{g^2(M^2)}{4\pi} =\frac{1}{\beta_2 \ln(M^2/\Lambda_{QCD}^2)} \, ,
\end{equation}
where $\beta_2=(11N_c-2N_f)/12\pi$ and $N_f$ is the number of quark flavors. We choose the infrared scale $m^2\sim \Lambda_{QCD}^2$. After substitution of $\mu$ from 
Eq. \eqref{mu_Qs}, replacing $g$ with $g(Q^2)$, and using the expression for the initial energy density $\varepsilon_0$ from Eq. \eqref{initial_energy}, Eq. \eqref{a2n} can be written as
\begin{equation}\label{a2n2}
\begin{split}
\mathcal{A}_{(2n)} &=\frac{2\varepsilon_0}{\ln(Q^2/m^2)} \left[\frac{4 Q_s^2\ln(Q_s^2/m^2)}{\ln(Q^2/m^2)}\right]^n \\
&\qquad \times \sum_{k=1}^{n} \mathsf{F}_{2n}(k) \left[\frac{Q^2\ln(Q^2/m^2)}{4Q^2_s\ln(Q^2_s/m^2)}\right]^k \\
& +\frac{2\varepsilon_0}{\ln^2(Q^2/m^2)} \left[\frac{4 Q_s^2\ln(Q_s^2/m^2)}{\ln(Q^2/m^2)}\right]^n  \\
&\qquad \times \sum_{k=2}^{n} \mathsf{G}_{2n}(k) \left[\frac{Q^2\ln(Q^2/m^2)}{4Q^2_s\ln(Q^2_s/m^2)}\right]^k\\
& + 2\varepsilon_0 \left[\frac{4 Q_s^2\ln(Q_s^2/m^2)}{\ln(Q^2/m^2)}\right]^n \mathsf{H}_{2n} \, .
\end{split}
\end{equation}
In obtaining Eq. \eqref{a2n2}, we included the running coupling constant expression Eq. \eqref{running_coupling} at two energy scales $\alpha_s(Q^2)$ and $\alpha_s(Q_s^2)$. The UV cut-off scale $Q$ sets the upper validity bound of the classical MV model which can not be determined by the model itself.  The hard partons with transverse momentum $p_{\perp}>Q$ can be described by perturbative QCD and are responsible for minijet producton, while the soft partons with transverse momentum  $p_{\perp}<Q$ can be effectively described by the classical fields and are responsible for the formation of the QGP.  Physical quantities like the energy density should be insensitive to the cut-off chosen after the soft processes are matched with the hard processes, as pointed out in \cite{Fries:2006pv}. On the other hand, the saturation scale $Q_s$ describes the typical transverse momentum scale of the saturated gluons. The scale is dynamically generated and depends on the collision energy and the size of the colliding nuclei. 

In the MV model, the large-x color sources are assumed to be uncorrelated on the transverse plane. This requires a finer resolution scale $Q$ beyond the typical coherent soft gluon scale $Q_s$. Therefore, we assume that $Q^2\gg Q_s^2$ \cite{Iancu:2003xm,Ferreiro:2001qy}. Then the summations in Eq. \eqref{a2n2} can be approximated by keeping only the leading $k=n$ terms and the last line of Eq. \eqref{a2n2} can be dropped.  Hence
\begin{equation}\label{a2n_Q}
\mathcal{A}_{(2n)} \simeq 2\varepsilon_0 \mathsf{F}_{2n}(n) \frac{Q^{2n}}{\ln (Q^2/m^2)} 
+2\varepsilon_0 \mathsf{G}_{2n}(n) \frac{Q^{2n}}{\ln^2 (Q^2/m^2)} \, .
\end{equation}
This is the expression for $\mathcal{A}_{(2n)}$ with which we work from now on. 

Before calculating the numerical coefficients $\mathsf{F}_{2n}(n)$ and $\mathsf{G}_{2n}(n)$, we would like to discuss the problem of including running coupling effects by replacing $\alpha_s$ with $\alpha_s(M^2)$. First of all, if we absorb the strong coupling constant into the vector potential $gA_{\mu} \rightarrow A_{\mu}$, the pure gauge field sector of the QCD lagrangian becomes $\mathcal{L}_G = -\frac{1}{4g^2} (F^a_{\mu\nu})^2$. Then the equations of motion Eq. \eqref{eoms} and the constraint equation Eq. \eqref{constraint} have no explicit dependence on the strong coupling constant $g$. That means the time evolution of the \textit{gauge fields} are free from the strong coupling constant and its running effects. Instead, the strong coupling constant is shuffled into the color charge sources $\rho_i \rightarrow g^2\rho_i$ $(i=1,2)$. As a result, the initial conditions Eq. \eqref{ics}, via their dependence on $A_1$ and $A_2$, are functions of the strong coupling constant. Moreover, the expression for the energy-momentum tensor would change correspondingly to $T^{\mu\nu}\rightarrow \frac{1}{g^2} T^{\mu\nu}$. Here comes the subtle point. When we consider the pressure isotropization in next section, the quantities we concentrate on are $p_T/\varepsilon$ and $p_L/\varepsilon$. The prefactor $1/g^2$ before $T^{\mu\nu}$ and the dependence on color charge sources $g^2\rho_i$, which are both reflected in the initial energy density $\varepsilon_0$, will cancel out. However, in arriving at Eq. \eqref{a2n2}, running coupling effects are encoded in the summations apart from the prefactor $\varepsilon_0$. As a consequence, we obtain the $Q^{2n}$ factor in Eq. \eqref{a2n_Q} under the leading $Q^2$ approximation. It will result in a $Q_s^{2n}$ factor in the expression for $\mathcal{A}_{(2n)}$ if no running coupling effects are included.  Summarizing, when studying pressure isotropization, the running coupling does affect the time evolution although its origin is in the initial conditions. 

Keeping in mind the places where the running coupling constant plays a role, replacing $\alpha_s$ with $\alpha_s(M^2)$ implies the inclusion of quantum fluctuation effects in these places. This is in principle beyond the scope of a pure classical model (the MV model). By the inclusion of a running coupling constant, we estimate the effects of quantum corrections in our classical description, specifically the initial dynamics at $\tau=0$. We expect interactions involving gluons as well as quarks due to their contributions in the running coupling constant expression Eq.\eqref{running_coupling}.  A taste of the problem can be found in the calculation of the QCD beta function using the background field method \cite{Peskin:1995ev} and the derivation of the JIMWLK equation using the Schwinger-Keldysh formalism \cite{Jeon:2013zga}.  Direct calculation of the gluonic Gaussian fluctuations on the classical fields have already been explored in \cite{Epelbaum:2013waa}. 
 
\subsection{Resummation}

Now we are ready to evaluate the coefficients $\mathsf{F}_{2n}(n)$ and $\mathsf{G}_{2n}(n)$. The leading $Q^2$ approximation to $\mathcal{A}_{(2n)}$ from Eq. \eqref{a2n_Q} is equivalent to keeping only the terms containing $2n$ derivatives in Eq. \eqref{a2n_cal}. As stated before, $D^i=D^i_1-igA^i_2=D^i_2-igA^i_1$, and terms in Eq. \eqref{a2n_cal} are organized according to the number of derivatives $D^i_1$ and $D^i_2$. To put it another way, the coupled nonlinear recursion solutions \eqref{recursive_A} become decoupled and linearized in the leading $Q^2$ approximation, namely
\begin{equation}
\begin{split}
&A^{\eta}_{(2n)} = \frac{1}{2n(2n+2)} \left[D^i_{(0)},\left[D^i_{(0)},A^{\eta}_{(2n-2)}\right]\right] \, , \\
&A^i_{(2n)} = \frac{1}{(2n)^2} \left[D^j_{(0)},F^{ji}_{(2n-2)}\right] \, . \\
\end{split}
\end{equation}
The components of the field strength tensor $F^{\mu\nu}$ are consequently independent of each other and are solved recursively.
\begin{equation}\label{general_solution}
\begin{split}
&\tilde{F}^{\tau\eta}_{(2n)} = -\frac{1}{[(2n)!!]^2} D^{\{2n\}} E_0 \, , \\
&B_{(2n)} =\frac{1}{[(2n)!!]^2} D^{\{2n\}} B_0 \, , \\
&\tilde{F}^{\eta i}_{(2n-1)} = \frac{2n}{[(2n)!!]^2}[D^i,D^{\{2n-2\}}E_0] \, , \\
&F^{\tau i}_{(2n-1)} = \frac{2n}{[(2n)!!]^2}\epsilon^{ij}[D^j,D^{\{2n-2\}}B_0] \, .\\
\end{split}
\end{equation}
Here we used $E_0 = \tilde{F}^{\tau\eta}_{(0)}$ and $B_0=-\frac{1}{2}\epsilon^{ij}F^{ij}_{(0)}$. The double factorial is $(2n)!!= 2n\times (2n-2) \times (2n-4) \ldots \times 2$. The $D^{\{2n\}}$ represents nested commutators of 2n folds
\begin{equation}
D^{\{2n\}} E_0 =[D^{i_n},[D^{i_n},\ldots [D^{i_1},[D^{i_1},E_0]]\ldots ]] \, .
\end{equation}
Substituting the above expressions into $\mathcal{A}_{(2n)}$ gives
\begin{equation}
\begin{split}
&\mathcal{A}_{(2n)} = \frac{1}{2} \sum_{k=0}^{n} \left(\tilde{F}^{\tau\eta}_{(2n-2k)}\tilde{F}^{\tau\eta}_{(2k)} +B_{(2n-2k)}B_{(2k)}\right)\\
& = \frac{1}{2}(-1)^nf_{(2n)} \left[D^{\{n\}}E_0D^{\{n\}}E_0 + D^{\{n\}}B_0D^{\{n\}}B_0 \right] \, , \\
\end{split}
\end{equation}
where $f_{(2n)} =\binom{2n}{n}/(2^n n!)^2$ contains the binomial coefficient $\binom{n}{k} =n!/(n-k)!k!$.
After taking the statistical average and carrying out contractions for both color and spatial indexes, we find
\begin{equation}
\begin{split}
&\mathsf{F}_{2n}(n) = \frac{(-1)^n}{n} f_{(2n)} \, , \quad n\geq 1\\
&\mathsf{G}_{2n}(n) = \frac{(-1)^n}{4} \mathcal{C}_{\pm}(2n)\, f_{(2n)} \, , \quad n\geq 2\\
\end{split}
\end{equation}

General expressions for $\mathcal{C}_{\pm}(2n)$ are given in Appendix \ref{c2n}.  (Here the $\pm$ refers to even or odd values of $n$.)  Unfortunately we were not able to find closed forms for them. Using the expression for $\mathcal{A}_{(2n)}$ in Eq. \eqref{a2n_Q}, we can sum all the terms in the expression for $\mathcal{A}$ in Eq. \eqref{aandb} to get
\begin{equation}
\begin{split}
\mathcal{A}= &\varepsilon_0 + \frac{2\varepsilon_0}{\ln (Q^2/m^2)} \sum_{n=1}^{\infty} \mathsf{F}_{2n}(n) (Q\tau)^{2n}\\
 &+ \frac{2\varepsilon_0}{\ln^2 (Q^2/m^2)} \sum_{n=2}^{\infty} \mathsf{G}_{2n}(n) (Q\tau)^{2n} \, .
\end{split}
\end{equation}
The first summation can be expressed in closed form as
\begin{equation}
\begin{split}
 &\sum_{n=1}^{\infty} \mathsf{F}_{2n}(n) (Q\tau)^{2n} = \sum_{n=1}^{\infty} \frac{(-1)^n}{n} f_{(2n)}(Q\tau)^{2n}\\
 &=-\frac{1}{2}(Q\tau)^2\left[{}_3F_4(1,1,\frac{3}{2};2,2,2,2;-(Q\tau)^2)\right] \, .
 \end{split}
\end{equation}
Here ${}_p F_q(a_1,\ldots,a_p;b_1,\dots,b_q;z)$ is the generalized hypergeometric function
\begin{equation}
{}_p F_q(a_1,\ldots,a_p;b_1,\dots,b_q;z)=\sum_{n=0}^{\infty}\frac{(a_1)_n\ldots(a_p)_n}{(b_1)_n\ldots(b_q)_n}\frac{z^n}{n!} \, .
\end{equation}
 However, the second summation does not have a closed form expression as far as we know. We will use $\mathsf{G}_A(Q\tau)$ to denote the second summation.
\begin{equation}
\begin{split}
&\mathsf{G}_A(Q\tau)=\sum_{n=2}^{\infty} \frac{(-1)^n}{4} \mathcal{C}_{\pm}(2n) f_{(2n)} (Q\tau)^{2n} \\
& = \frac{1}{4}\left[8f_{(4)}(Q\tau)^4-18f_{(6)}(Q\tau)^6+ \frac{118}{3}f_{(8)}(Q\tau)^8+\ldots\right]
\end{split}
\end{equation}
Then we get our final expression for $\mathcal{A}$
\begin{equation}\label{resumA}
\begin{split}
\mathcal{A} &= \varepsilon_0  + \frac{2\varepsilon_0}{\ln^2 (Q^2/m^2)} \mathsf{G}_A(Q\tau) \\
&- \frac{\varepsilon_0}{\ln (Q^2/m^2)} (Q\tau)^2 \left[{}_3F_4(1,1,\frac{3}{2};2,2,2,2;-(Q\tau)^2)\right] \, .
\end{split}
\end{equation} 
With the relation \eqref{aandb2} between $\mathcal{A}$ and $\mathcal{B}$ we immediately obtain
\begin{equation} \label{resumB}
\begin{split}
\mathcal{B} =& \frac{2\varepsilon_0}{\ln (Q^2/m^2)} \left(1-[J_0(Q\tau)]^2-[J_1(Q\tau)]^2\right)\\
&+ \frac{2\varepsilon_0}{\ln^2 (Q^2/m^2)} \mathsf{G}_B(Q\tau)\\
\end{split}
\end{equation}
where $J_n(x)$ is the Bessel function of the first kind and 
\begin{equation}
\begin{split}
&\mathsf{G}_B(Q\tau)=-\sum_{n=2}^{\infty}\frac{(-1)^n}{4} \frac{2n}{2n+2} C_{\pm}(2n)f_{(2n)}(Q\tau)^{2n}\\
&=  -\left[\frac{4}{3}f_{(4)}(Q\tau)^4-\frac{27}{8}f_{(6)}(Q\tau)^6+ \frac{118}{15}f_{(8)}(Q\tau)^8+\ldots\right] \, .
\end{split}
\end{equation}
Equations \eqref{resumA} and \eqref{resumB} are the main results in this paper. They both contain two parts: one part can be expressed in closed form while the other part can only be expressed as an infinite power series. 

\section{Numerical Results} \label{numerical}

In the central rapidity region where $\eta =0$, the energy density and pressures from Eq. \eqref{emtensor2} are expressed as $\varepsilon = \mathcal{A}+\mathcal{B}$, $P_T = \mathcal{A}$ and $P_L = - \mathcal{A}+\mathcal{B}$. With the results for $\mathcal{A}$ in Eq. \eqref{resumA} and $\mathcal{B}$ in Eq. \eqref{resumB}, we are ready to explore the time evolution of energy density and pressures. As functions of $Q\tau$, the only unknown paramter in $\mathcal{A}$ and $\mathcal{B}$ is $a\equiv\ln Q^2/m^2$ because the initial energy density $\varepsilon_0$ acts as an overall prefactor which can be rescaled out as long as we are not interested in the absolute values. The value of $a$ has to be large as we assumed $Q^2\gg m^2$, and we also work in the high-momentum regime where $Q^2\gg Q_s^2$. In the following numerical calculations we choose $a=6$, then 
$Q=4.0$ GeV when $m=\Lambda_{QCD}=0.2$ GeV, and $Q_s\sim 1.2$ GeV. This will make $Q^2/m^2 \sim 400$ and $Q^2/Q_s^2 \sim 12$, which are consistent with our assumptions. We rescale energy density and pressures as $\varepsilon/\varepsilon_0$, $P_T/\varepsilon_0$, $P_L/\varepsilon_0$. 

\begin{figure}
\includegraphics[scale = 0.62]{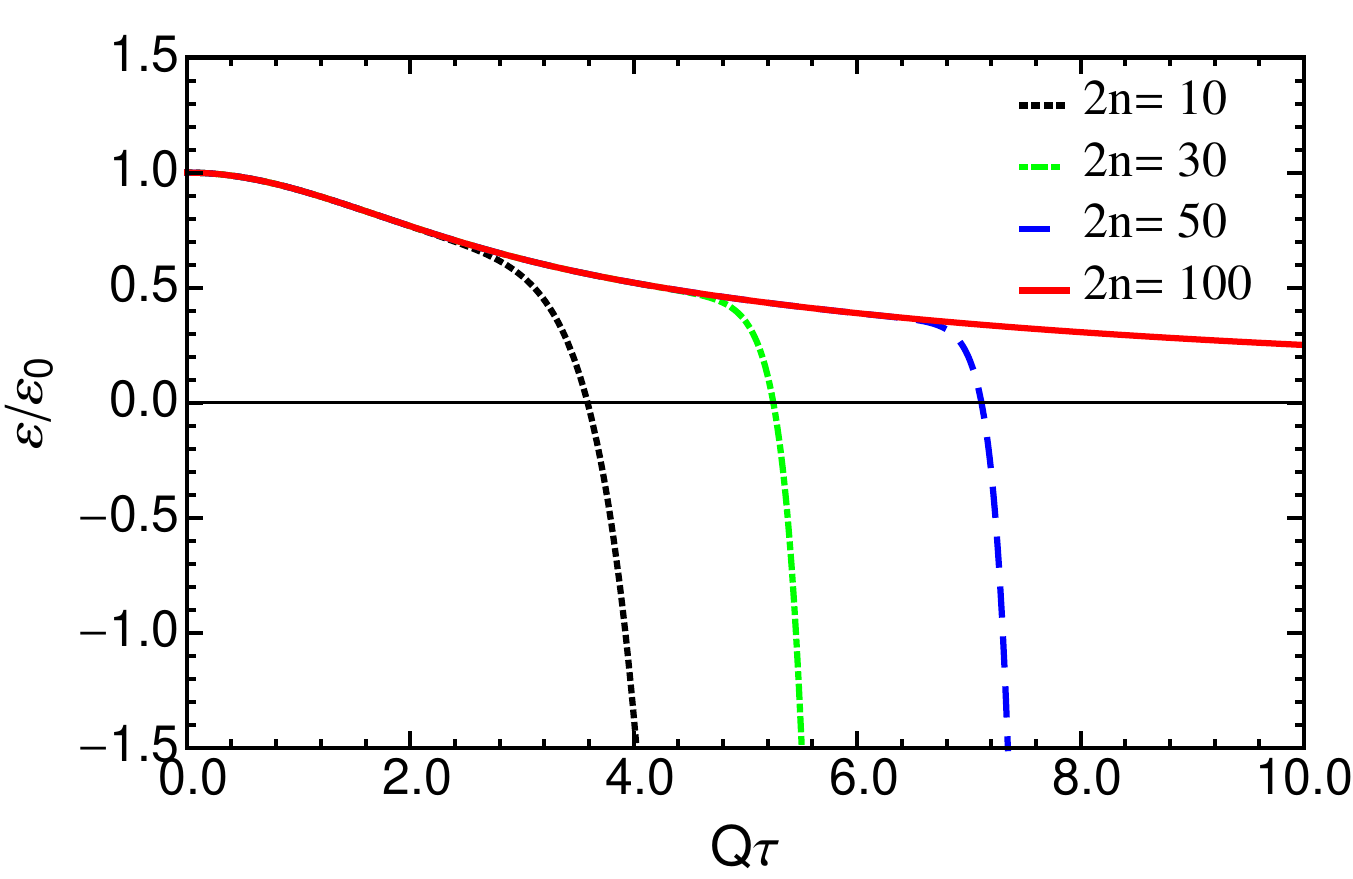}
\caption{(color online) Energy density $\varepsilon/\varepsilon_0$ as a function of $Q\tau$ with different calculated accuracy. The black dotted line, green dot-dashed line, blue dashed line and the red solid line represent approximations of $\mathsf{G}_A(Q\tau)$ and $\mathsf{G}_B(Q\tau)$ to the order of $(Q\tau)^{10}$, $(Q\tau)^{30}$, $(Q\tau)^{50}$, and $(Q\tau)^{100}$, respectively.}
\label{fig:energydensity}
\end{figure}

\begin{figure}
\includegraphics[scale = 0.62]{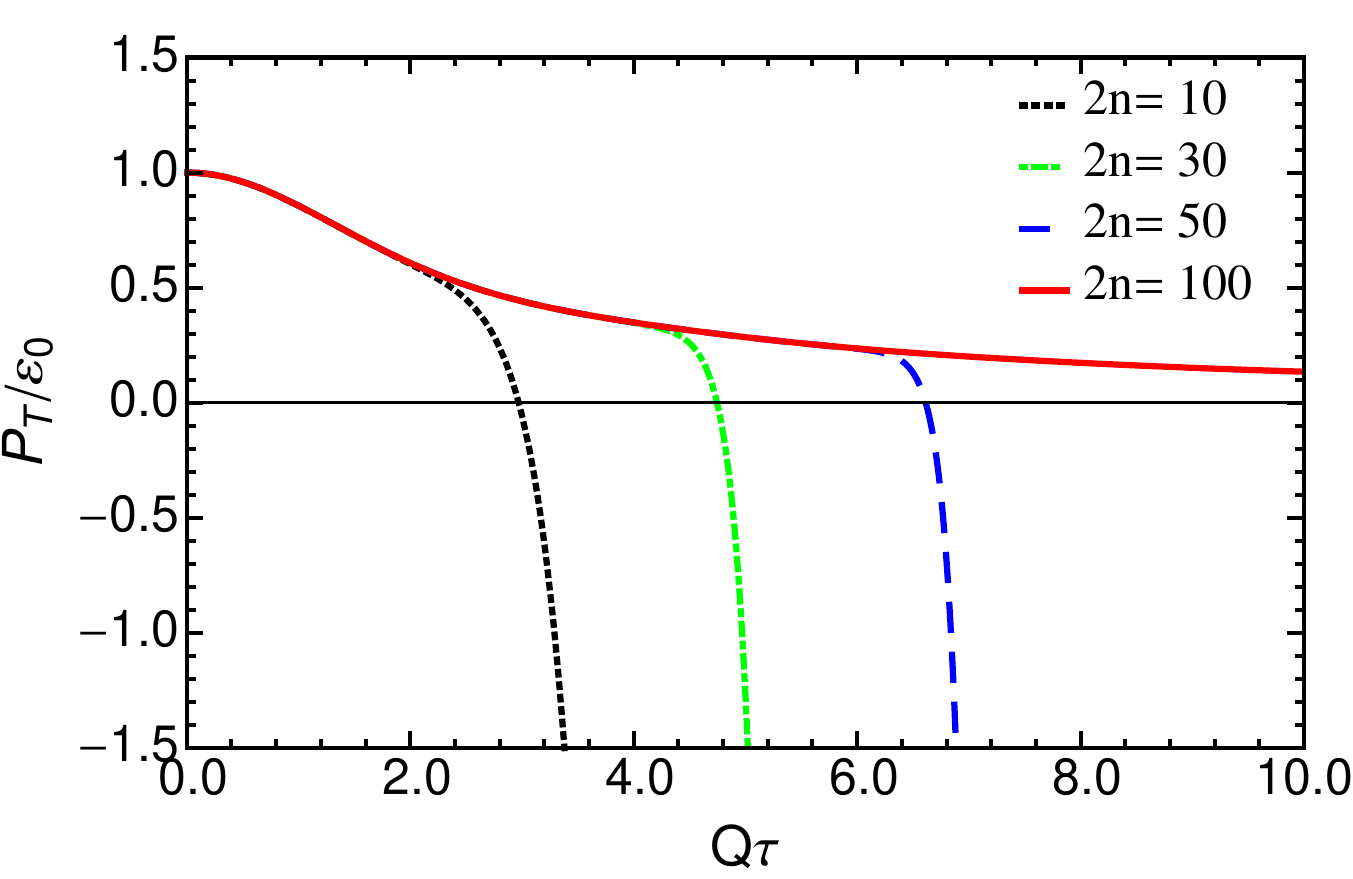}
\caption{(color online) Transverse pressure $P_T/\varepsilon_0$ as a function of $Q\tau$ with different calculated accuracy.  The notation is the same as in Fig. \ref{fig:energydensity}.}
\label{fig:pressureT}
\end{figure}

In Fig. \ref{fig:energydensity} the time evolution of the energy density is shown with different calculated accuracy. We approximate $\mathsf{G}_A(Q\tau)$ and $\mathsf{G}_B(Q\tau)$ in Eqs.  \eqref{resumA} and \eqref{resumB} up to the order of $(Q\tau)^{10}$, $(Q\tau)^{30}$, $(Q\tau)^{50}$, and $(Q\tau)^{100}$. The energy density decreases with time and sharply drops to negative values for a given order. Negative energy density is unphysical, we can only trust the time evolution of the energy density in the region where $\varepsilon>0$. Moreover, increasing the order of accuracy makes the value of $Q\tau$ (when $\varepsilon=0$) larger, thus enlarging the time interval for the validity of the energy density. As we can see, the result at the order $(Q\tau)^{100}$ is sufficient to describe the time evolution within the interval $0\leq Q\tau \leq10$ chosen. Additional higher order terms will not change the behavior of the energy density in this interval. Results are only shown within this interval for two reasons. Late time evolution of the glasma is dominated by quantum effects, which makes the prediction of the pure classical description questionable.  Also, large values of $Q \tau$ in the power series expressions introduce large round-off errors in numerical precision which can result in unphysical behavior.  Similar behavior is found in Fig. \ref{fig:pressureT} for the transverse pressure $P_T$.
 
\begin{figure}
\includegraphics[scale=0.62]{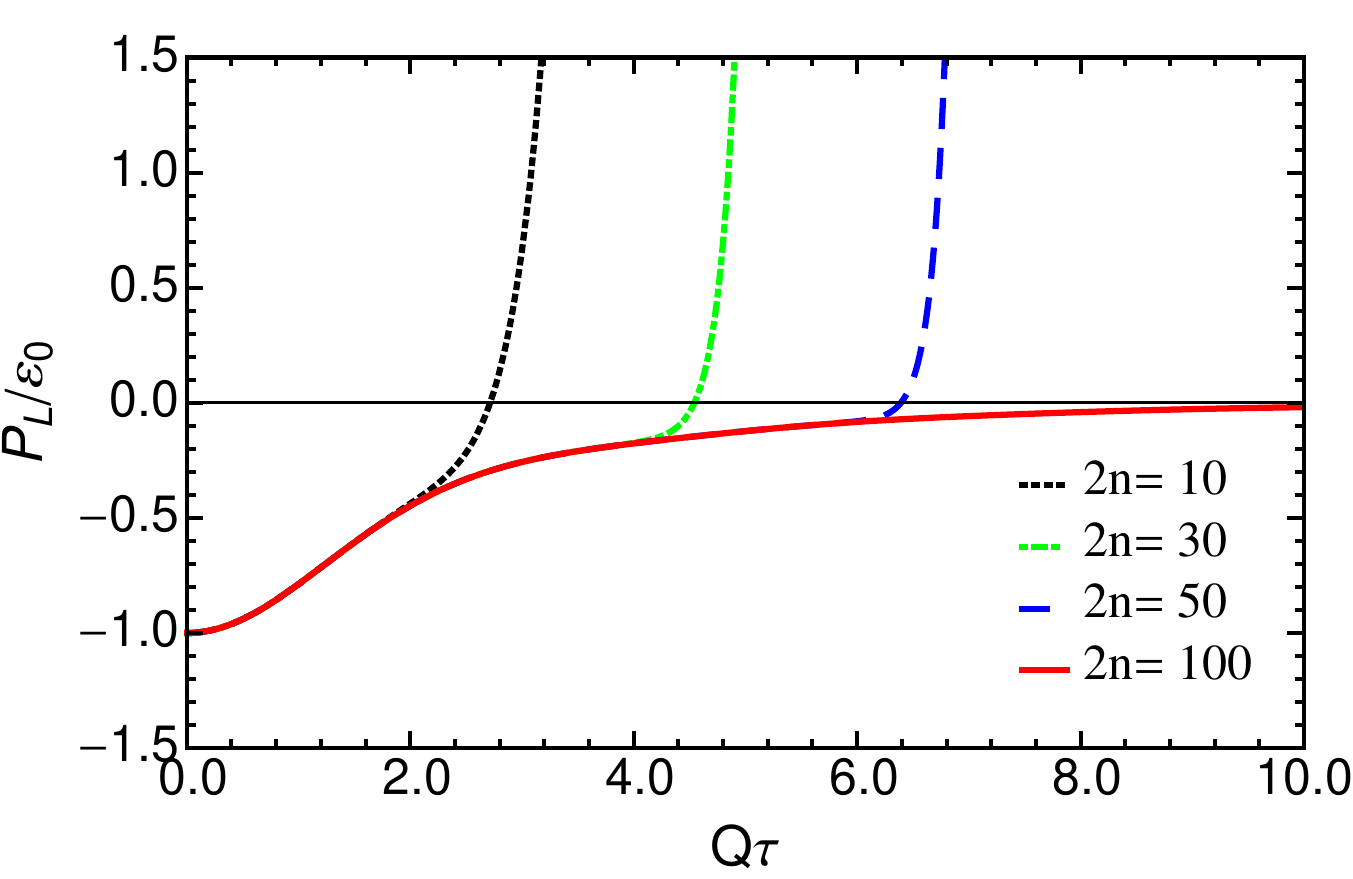}
\caption{(color online) Longitudinal pressure $P_L/\varepsilon_0$ as a function of $Q\tau$ with different calculated accuracy.  The notation is the same as in Fig. \ref{fig:energydensity}.}
\label{fig:pressureL}
\end{figure}

Fig. \ref{fig:pressureL} shows the rescaled longitudinal pressure $P_L/\varepsilon_0$ evolution. The initial negative value of $P_L$ originates from the longitudinal motion of the two nuclei and the back-reaction of the glasma on the receding nuclei.  The magnitude of the longitudinal pressure decreases as the nuclei recede from each other. From the results at the order of $2n=100$, one can see the longitudinal pressure is still negative but tending towards zero. This implies that the classical gluon field description cannot realize pressure isotropization which requires the longitudinal pressure to become positive.

\begin{figure}
\includegraphics[scale = 0.75]{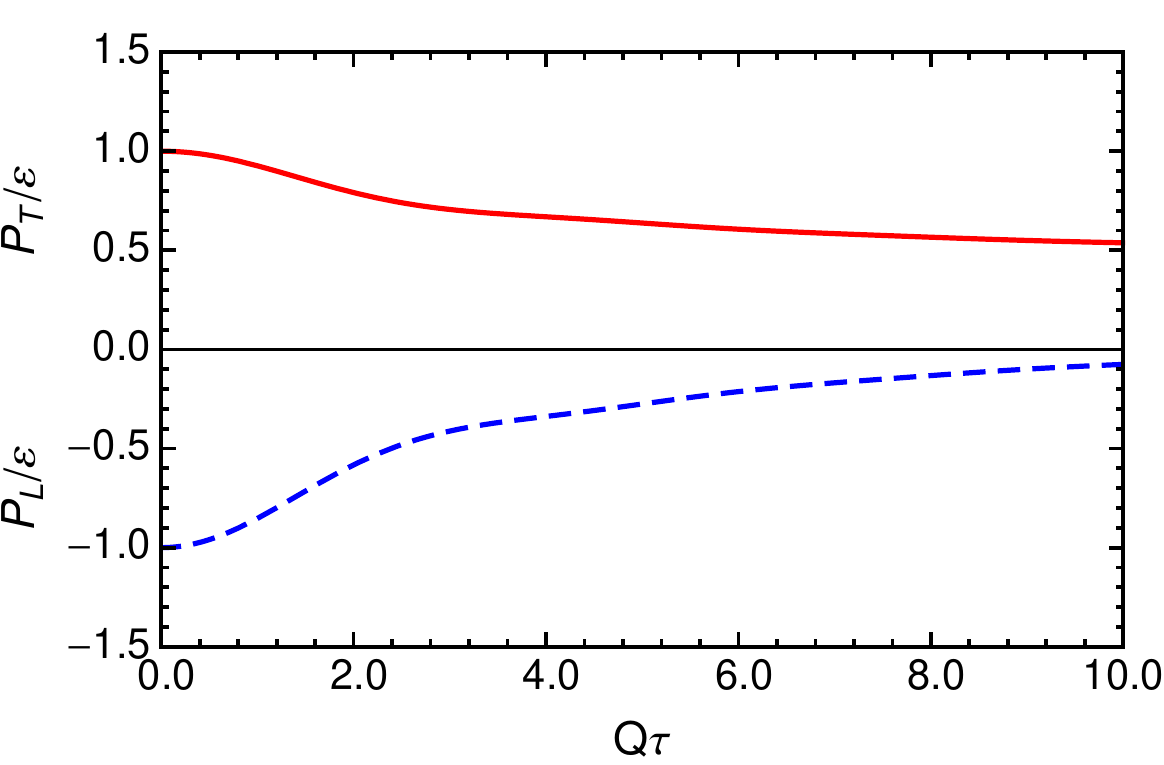}
\caption{ (color online) The pressures $P_T/\varepsilon$ and $P_L/\varepsilon$ as functions of $Q\tau$ with different calculated accuracy. The lines in the upper half plane represent $P_T/\varepsilon$ while lines on the lower half plane represent $P_L/\varepsilon$.  The notation is the same as in Fig. \ref{fig:energydensity}. }
\label{fig:ptpl_e}
\end{figure}

In Fig. \ref{fig:ptpl_e} the time evolution of $P_T/\varepsilon$ and $P_L/\varepsilon$ are shown where $\mathsf{G}_A(Q\tau)$ and $\mathsf{G}_B(Q\tau)$ are truncated at the order of $2n=100$.  The ratio of the longitudinal and transverse pressures to the energy density are $P_T/\varepsilon \simeq 0.5$ and $P_L/\varepsilon \simeq 0$ at the time $Q\tau=10.0$ ($\tau =0.5\, \rm{fm/c}$). The results obtained here are identical to those of \cite{Chen:2015wia} when all terms are truncated at order $(Q\tau)^4$ and the input parameter $a$ is chosen appropriately. They are essentially the same as obtained in the classical SU(2) simulations in \cite{Gelis:2013rba}.  (There are small oscillations in their results which may be attributable  to the finite lattice spacing or to the finite rapidity region of the space-time which was sampled and averaged over.  Our solution assumes boost invariance.)  When they include initial quantum fluctuations, and using a value of $g=0.5$, they find a positive longitudinal pressure but still a sizeable anisotropy of $P_L/P_T \sim 0.6$ at a time of 1 fm/c.  For a smaller coupling of $g=0.1$ the initial quantum fluctuations have very little effect on the classical solution.

\section{Conclusion} 
\label{conclusion}

In this paper we calculated analytical expressions for the energy-momentum tensor of the glasma created in high energy heavy ion collisions to all orders in the leading $Q^2$ approximation with the inclusion of running coupling effects. These expressions are functions of the dimensionless quantity $Q\tau$.  They contain one part that can be expressed in closed form and another part that can only be expressed as an infinite power series with all the coefficients under control. Numerical calculations suggest that finite order results ($2n\sim 100$) are sufficient to describe the time evolution of the energy density and pressures within the time interval $0\leq Q\tau \leq 10$.  We found that the classical gluon field description predicts pressure anisotropy up to the time $\tau = 0.5\, \rm{fm/c}$ (using $Q \tau = 10$ and $Q = 4$ GeV). 

There are obviously a number of issues that require further investigation.  First of all, while the leading $Q^2$ approximation is self-consistent, we ignored all subleading terms. Their contributions to the early time evolution should be addressed to have a more robust prediction.  Second, our calculations assumed slab on slab collisions.  Including variation of the initial color charge densities is necessary to quantitatively understand the early time transverse flow effects \cite{Chen:2015wia}.  Third, our analysis of the effective classical Yang-Mills theory includes quantum corrections by replacing the strong coupling constant $\alpha_s$ by $\alpha_s(M^2)$. More detailed study of quantum effects helping to drive the system towards isotropization are needed.  Fourth, plasma instabilities due to initial momentum anisotropy are not captured in our small-$\tau$ power series solution after ensemble averaging the initial state; they would need to be tracked separately.  Both quantum effects and plasma instabilities may play a role in thermalization.  Fifth, studies have suggested a universal attractor solution which governs the late time evolution in the classical regime \cite{Berges:1,Berges:2,Berges:3,Berges:4,Berges:5}.  Finally, it would be interesting to perform something akin to a sudden approximation where the classical fields are converted to partons with subsequent evolution of the system described by a Boltzmann equation \cite{Kurkela:2015}. 

\section*{Acknowledgement}

We thank Michal Heller for noticing an error in a {\it Mathematica} program used in an earlier version of this paper.  This work was supported by the U. S. Department of Energy grant  DE-FG02-87ER40328.

\begin{appendix}

\section{Gluon Correlation Functions}\label{appendixA}

The method of calculating higher twist gluon correlation functions is described in Ref. \cite{Chen:2015wia}. We sketch the main steps here. All the correlation functions are expressed in terms of gradients of $\gamma(\vec{x}_{\perp},\vec{y}_{\perp})$ which is
\begin{equation}
\gamma(\vec{x}_{\perp},\vec{y}_{\perp})= \mu \int\frac{d^2\vec{k}_{\perp}}{(2\pi)^2} e^{i\vec{k}_{\perp} (\vec{x}_{\perp} - \vec{y}_{\perp})} \frac{1}{(k_{\perp}^2+m^2)^2} \, .
\end{equation}
A few examples are
\begin{equation}
\langle A^i_a(\vec{x}_{\perp}) A^j_b(\vec{x}_{\perp}) \rangle =\delta_{ab}\frac{g^2}{d_A} \nabla^i_x\nabla^j_y\gamma(\vec{x}_{\perp},\vec{y}_{\perp}) \vert_{\vec{y}_{\perp}\rightarrow \vec{x}_{\perp}} \, ,
\end{equation}
\begin{equation}
\begin{split}
&\langle (D^kA^i)_a(\vec{x}_{\perp}) (D^lA^j)_b(\vec{x}_{\perp}) \rangle \\ &=\delta_{ab}\frac{g^2}{d_A}\times \nabla^k_x\nabla^l_y\nabla^i_x\nabla^j_y\gamma(\vec{x}_{\perp},\vec{y}_{\perp}) \vert_{\vec{y}_{\perp}\rightarrow \vec{x}_{\perp}} \, ,\\
\end{split}
\end{equation}
and
\begin{equation}
\begin{split}
&\langle (D^kD^lA^i)_a(\vec{x}_{\perp}) (D^mD^nA^j)_b(\vec{x}_{\perp}) \rangle\\ &=\delta_{ab}\frac{g^2}{d_A} \nabla^k_x\nabla^l_x\nabla^m_y\nabla^n_y\nabla^i_x\nabla^j_y\gamma(\vec{x}_{\perp},\vec{y}_{\perp}) \vert_{\vec{y}_{\perp}\rightarrow \vec{x}_{\perp}}\, .
\end{split}
\end{equation}
We only consider terms containing even numbers of covariant derivatives, because terms with odd numbers of covariant derivatives vanish by homogeneity and isotropy.  This can be seen from explicit calculations like
\begin{equation}\label{uv_calculate}
\begin{split}
&\nabla^k_{x,y}\nabla^l_{x,y}\nabla^i_x\nabla^j_y\gamma(\vec{x}_{\perp},\vec{y}_{\perp}) \vert_{\vec{y}_{\perp}\rightarrow \vec{x}_{\perp}}\\
&=-\mu\int \frac{d^2\vec{k}_{\perp}}{(2\pi)^2} e^{i\vec{k}_{\perp}(\vec{x}_{\perp}-\vec{y}_{\perp})} \frac{k^k_{\perp}k^l_{\perp}k^i_{\perp}k^j_{\perp}}{(k_{\perp}^2+m^2)^2} \\
&=-\mu\int\frac{d^2\vec{k}_{\perp}}{(2\pi)^2}\frac{k_{\perp}^4}{(k_{\perp}^2+m^2)^2} \frac{1}{8}(\delta^{kl}\delta^{ij}+\delta^{ki}\delta^{lj}+\delta^{kj}\delta^{li})\\
&=-\frac{\mu}{32\pi} \int^{Q^2} dk^2_{\perp} \frac{k^4_{\perp}}{k^4_{\perp}}(\delta^{kl}\delta^{ij}+\delta^{ki}\delta^{lj}+\delta^{kj}\delta^{li})\\
&=-\frac{\mu}{32\pi} Q^2 (\delta^{kl}\delta^{ij}+\delta^{ki}\delta^{lj}+\delta^{kj}\delta^{li}) \, .
\end{split}
\end{equation}
In the above integration, we only kept the leading $Q^2$ terms due to the assumption that $Q^2\gg m^2$. Higher twist correlation functions have more spatial indexes to deal with, for example
\begin{equation}
\begin{split}
&\nabla^k_x\nabla^l_x\nabla^m_y\nabla^n_y\nabla^i_x\nabla^j_y\gamma(\vec{x}_{\perp},\vec{y}_{\perp}) \vert_{\vec{y}_{\perp}\rightarrow \vec{x}_{\perp}}\\
&=\mu\int \frac{d^2\vec{k}_{\perp}}{(2\pi)^2} e^{i\vec{k}_{\perp}(\vec{x}_{\perp}-\vec{y}_{\perp})} \frac{k^k_{\perp}k^l_{\perp}k^m_{\perp}k^n_{\perp}k^i_{\perp}k^j_{\perp}}{(k_{\perp}^2+m^2)^2} \\
&=\frac{\mu}{4\pi} \frac{Q^4}{2} \frac{1}{48} \Delta^{klmnij} \, ,
\end{split}
\end{equation}
where the tensor $\Delta^{klmnij}$ is defined below.  The momentum indices can be grouped as
\begin{equation}
\begin{split}
&k^k_{\perp}k^l_{\perp}k^m_{\perp}k^n_{\perp}k^i_{\perp}k^j_{\perp}\\
&=\frac{k_{\perp}^6}{48} \Big(\delta^{ij}\delta^{kl}\delta^{mn} +\delta^{ij}\delta^{km}\delta^{ln}+\delta^{ij}\delta^{kn}\delta^{lm}\\
&+\delta^{ik}\delta^{jl}\delta^{mn}+\delta^{ik}\delta^{jm}\delta^{ln}+\delta^{ik}\delta^{jn}\delta^{lm}\\
&+\delta^{il}\delta^{jk}\delta^{mn}+\delta^{il}\delta^{jm}\delta^{kn}+\delta^{il}\delta^{jn}\delta^{km}\\
&+\delta^{im}\delta^{jk}\delta^{ln}+\delta^{im}\delta^{jl}\delta^{kn}+\delta^{im}\delta^{jn}\delta^{lk}\\
&+\delta^{in}\delta^{jk}\delta^{lm}+\delta^{in}\delta^{jl}\delta^{km}+\delta^{in}\delta^{jm}\delta^{lk}\Big) \, .
\end{split}
\end{equation}
Therefore, we have to address the problem of complicated spatial index contractions during the resummation. We define $\Delta^{i_1i_2\ldots i_{2n}}$ as the summation of all possible products of Kronecker delta functions with spatial indexes $i_1,i_2,\ldots,i_{2n}$. A few examples are
\begin{equation}
\begin{split}
&\Delta^{mp} \equiv \delta^{mp} \, , \\
&\Delta^{ijmp} \equiv \delta^{ij}\Delta^{mp}+\delta^{im}\Delta^{jp}+\delta^{ip}\Delta^{jm} \, , \\
&\Delta^{klijmp} \equiv \delta^{kl}\Delta^{ijmp}+\delta^{ki}\Delta^{ljmp}+\delta^{kj}\Delta^{limp}\\
&\qquad\qquad\quad +\delta^{km}\Delta^{lijp}+\delta^{kp}\Delta^{lijm} \, .
\end{split}
\end{equation}
In general, the normalized expression is
\begin{equation}
\frac{1}{(2n)!!}\Delta^{i_1i_2\ldots i_{2n}}
\end{equation}
Notice that $\Delta^{i_1i_2\ldots i_{2n}}$ is totally symmetric under exchange of any two indexes. After explicit calculation, we obtain the following contraction identity which is used throughout the resummation process.
\begin{equation}
\begin{split}
&(\delta^{mn}\delta^{pq}+\epsilon^{mn}\epsilon^{pq}) \frac{1}{(2n)!!}\Delta^{i_1i_2\ldots i_{2n-2} mp}\frac{1}{(2n)!!}\Delta^{i_1i_2\ldots i_{2n-2} nq}\\ 
&=\frac{(2n-2)!}{[(2n-2)!!]^2} \\
\end{split}
\end{equation}


\section{$\mathcal{C}_{\pm}(2n)$} \label{c2n}


The expressions for $\mathcal{C}_{\pm}(2n)$ are complicated. For $n$ an even integer $(n\geq 2)$
\begin{widetext}
\begin{equation}\label{cp2n}
\begin{split}
\mathcal{C}_+(2n)&=2 \, \bigg[\sum_{k=0}^{n/2}\sum_{l=0}^{n/2}\sum_{i=0}^{k}\sum_{j=0}^{l}\frac{1}{n-k-l}\frac{1}{k+l} \binom{n/2}{k+i}\binom{k+i}{2i}\binom{n/2}{l+j}\binom{l+j}{2j}\binom{2i+2j}{i+j}\\
&+\sum_{k=0}^{n/2-1}\sum_{l=0}^{n/2-1}\sum_{i=0}^{k}\sum_{j=0}^l\frac{1}{n-k-l-1}\frac{1}{k+l+1}\binom{n/2}{k+i+1}\binom{k+i+1}{2i+1}\binom{n/2}{l+j+1}\\
&\qquad\times \binom{l+j+1}{2j+1}\binom{2i+2j+2}{i+j+1} \bigg] \, .
\end{split}
\end{equation}
In the first line the term with $k=l=0$ is excluded, and the term with $k=l=n/2$ is excluded.
For n an odd integer $(n\geq 3)$
\begin{equation}\label{cm2n}
\begin{split}
\mathcal{C}_{-}(2n) =4 \, \bigg[ &\sum_{k=0}^{(n-1)/2}\sum_{l=0}^{(n-1)/2} \left(\frac{1}{n-k-l-1}\right)\left(\frac{1}{k+l+1}\right)\\
&\times \bigg[\sum_{i=0}^{k}\sum_{j=0}^{l} \binom{(n-1)/2}{k+i}\binom{k+i}{2i}\binom{(n-1)/2}{l+j}\binom{l+j}{2j}\binom{2i+2j}{i+j} \\
&+ \frac{1}{2}\binom{(n-1)/2}{k+i}\binom{k+i}{2i}\binom{(n-1)/2}{l+j+1}\binom{l+j+1}{2j+1} \binom{2i+2j+2}{i+j+1} \\
&+ \frac{1}{2}\binom{(n-1)/2}{k+i+1}\binom{k+i+1}{2i+1}\binom{(n-1)/2}{l+j}\binom{l+j}{2j} \binom{2i+2j+2}{i+j+1}\\
&+ \binom{(n-1)/2}{k+i+1}\binom{k+i+1}{2i+1}\binom{(n-1)/2}{l+j+1}\binom{l+j+1}{2j+1} \binom{2i+2j+2}{i+j+1} \bigg]\bigg] \, . 
\end{split}
\end{equation}
Also, the term with $k=l=(n-1)/2$ is excluded.
\end{widetext}

To understand the general structure of $\mathcal{C}_{+}(2n)$ and $\mathcal{C}_{-}(2n)$, consider the first line of expression \eqref{cp2n} as an example. At each order $2n$, we have $n$ derivatives $D^i$ to distribute between $A^m_1$ and $A^n_2$, and another $n$ derivatives to distribute between $A^p_1$ and $A^q_2$. The number of derivatives acting on $A_1^m$ and $A^p_1$ in total has to be even, otherwise their statistical averages will vanish. The same is true for $A_2^n$ and $A_2^q$. We have $n-2k$ derivatives $D^i_1$ acting on $A^m_1$ and $n-2l$ derivatives $D^i_1$ acting on $A^p_1$. But within the $n-2k$ (or $n-2l$) derivatives, we can choose either $D^i_1D^i_1$ or $D^i_1D^j_1$, which is why there are additional summation indexes $i$ or $j$. The two prefactors $1/(n-k-l)$ and $1/k+l$ come from the momentum space integral $\int dp^2 (p^2)^{n-k-l}$ and $\int dp^2 (p^2)^{k+l}$ when evaluating correlation functions. Finally, the binomial coefficients $\binom{2i+2j}{i+j}$ are due to spatial index contractions. 

\end{appendix}

\newpage

\bibliography{pressure_isotropization}
\end{document}